# *DAY TRADE* – **Do outro lado das estatísticas** [*]

*Roberto Ernani Porcher Junior* [**]

08 de dezembro de 2019

## RESUMO

O presente trabalho coloca em questionamento algumas ideias atualmente difundidas sobre a prática de operações específicas do mercado de capitais – as chamadas operações *day trade*. Partindo de proposições teóricas, o texto avança a uma análise detalhada do estudo intitulado "É possível viver de *day-trading?*" (CHAGUE e GIOVANNETTI, 2019), ao qual oferece contraponto. Essa investigação revela a existência de importantes elementos que ainda não estão devidamente sopesados no tratamento do tema em voga, acarretando perda de dimensões que são – ou deveriam ser – indissociáveis desse tipo de pesquisa. A conclusão alcançada é de que as evidências científicas até então existentes não demonstram que a adoção do *day trade* como profissão seja insustentável economicamente, nem comprovam a impossibilidade de evolução da performance operacional dos *day traders* ao longo do tempo.

Palavras-chave: *Day trade.* Investimentos. Especulação. Bolsa de valores. Mercado de capitais.

---

[**] Bacharel em Direito pela Pontifícia Universidade Católica do Rio Grande do Sul – PUCRS. Especialista em Finanças, Investimentos e Banking pela Pontifícia Universidade Católica do Rio Grande do Sul – PUCRS. Advogado em Porto Alegre/RS. Telefone: (+55) 51 98250-3333. Correio eletrônico: robertoporcher@gmail.com

**ABSTRACT**

This paper questions some current ideas about the practice of specific capital market operations - the so-called day trading operations. The text advanced from theoretical propositions to a detailed analysis of the study entitled "Is it possible to live by day-trading?" (CHAGUE and GIOVANNETTI, 2019), to which it offers a counterpoint. This investigation reveal the existence of important elements that are not yet properly weighed in the treatment of the current theme, leading to loss of dimensions that are - or should be - inseparable from this type of research. The conclusion reached is that the existing scientific evidence does not show that the adoption of the day trade as an occupation is economically unsustainable, nor does it prove the impossibility of evolution of the day traders' operational performance over time.

Key-words: Day trade. Investments. Speculation. Stock exchange. Capital market.

**1 – Introdução**

O presente artigo tem o espírito de contribuir para o alargamento da visão habitualmente lançada sobre a prática de operações específicas do mercado de capitais – as chamadas operações *day trade*. Partindo de proposições teóricas, mas considerando também os dados empíricos constantes de pesquisas existentes, o texto se constrói através do questionamento de ideias atualmente difundidas sobre a realidade dos *day traders* e apresenta concepções alternativas ou complementares para o enfrentamento deste tema.

Como pauta para desenvolvimento do texto está o estudo intitulado "É possível viver de *day-trading?*" (CHAGUE e GIOVANNETTI, 2019),[1] ao qual se oferece contraponto. No referido *paper*, elaborado a pedido da Comissão de Valores Mobiliários – CVM, foi retratada a dificuldade que pessoas físicas encontram para obter lucro significativo realizando operações *day trade*, e a conclusão veiculada foi de que não faz sentido adotar tal atividade como pretensa fonte de renda para a subsistência, pois os prejuízos superam os ganhos, e o desempenho piora ao longo do tempo. Em oposição a este veredito, nas próximas páginas vão problematizados alguns tópicos presentes no âmbito daquela pesquisa, sendo apresentados fundamentos que questionam e relativizam a universalização das suas conclusões.

O diálogo aqui proposto não se aprofunda na seara da expertise estatística, nem adentra ao campo das fórmulas matemáticas de escrutínio dos dados, dando ênfase, primordialmente, a reflexões sobre conceitos e questões práticas da atividade de *day trade*. Tal análise busca revelar a existência de importantes elementos que ainda não estão devidamente sopesados no tratamento científico do tema, acarretando perda de dimensões que são – ou deveriam ser – indissociáveis do objeto da pesquisa.

Sob o viés de desconstrução das evidências empíricas atualmente difundidas, em especial aquelas que sugerem ser insustentável economicamente a adoção da profissão de *day trade* por pessoas físicas, nasce oportunidade para fomentar e incentivar o aperfeiçoamento dos investidores interessados em se dedicar a essa atividade. Alinhada com este horizonte, está proposta uma investigação teórica onde se busca analisar a possibilidade de melhora no desempenho do *day trader* ao longo da sua experiência, como resultado de efetiva evolução ordenada da sua performance.

---

[1] Disponível em: https://cointimes.com.br/wp-content/uploads/2019/03/Viver-de-day-trading-1.pdf.

## 2 – Proposições teóricas

De início, cabe conceituar o *day trade* como sendo um tipo de negociação, geralmente ocorrida no ambiente de bolsa de valores, em que alguém abre uma posição, comprando ou vendendo determinada quantidade de um ativo/derivativo e, ainda no mesmo dia, fecha essa posição, revendendo ou recomprando a mesma quantidade do mesmo ativo/derivativo, de modo a retornar ao estado anterior à negociação. Assim, a operação *day trade* começa e termina no mesmo pregão, tendo como vida útil o período intradiário, e resultando em lucro ou prejuízo pela diferença *(spread)* entre o valor de abertura e de fechamento da posição.

Embora seja totalmente lícita, a prática do *day trade* é extremamente estigmatizada, o que pode ser explicado, em parte, por influência de aspectos culturais. Ensinamentos ministrados[2] por Pondé (2018) e Karnal (2018) demonstram a arqueologia de concepções enraizadas em diferentes estratos sociais, a exemplo da censura ao lucro voltado à concentração de capital, a qual tem matrizes ideológicas e religiosas que remontam séculos. Em países que se desenvolvem marcados por desigualdades, essa concepção cultural bem pode fornecer confortável suporte a um sentido de estranhamento ocasionado quando algumas pessoas auferem significativos valores de forma muito rápida. Tal performance também contraria o arquétipo mitológico de que o indivíduo só é merecedor de uma recompensa após penar longa e arduamente fazendo tarefas hercúleas, parecendo faltar legitimidade à riqueza gerada de forma acelerada.

Exemplo dessa carga valorativa presente no substrato sociocultural é a dicotomia existente entre os termos "investir" e "especular", sendo ao primeiro emprestada mais respeitabilidade do que ao segundo, este empregado em sentido pejorativo e preconceituoso. É como se "investir" tivesse bases sólidas e seguras, e "especular" fosse arriscadíssimo e aleatório; como se "investir" fosse galgar a longo prazo, e "especular" caracterizasse uma jogada para encurtar o caminho; como se "investir" significasse construir rentabilidade, e "especular" significasse buscar vantagem. Em verdade, é comum encontrar essa mesma distinção no ideário de grandes investidores, como Warren Buffett.[3]

---

[2] Algumas das citações constantes do texto refletem ensinamentos proferidos oralmente ao longo das disciplinas do curso de Especialização em Finanças, Investimentos e Banking, do Programa de Pós-Graduação da Escola de Negócios da Pontifícia Universidade Católica do Rio Grande do Sul – PUCRS, durante o ano de 2018.
[3] D'ÁVILA, Mariana. **Você sabe a diferença entre investir e especular? Warren Buffett explica**. InfoMoney. 2016. Publicação em site de conteúdo e notícias. Disponível em: https://www.infomoney.com.br/onde-investir/voce-sabe-a-diferenca-entre-investir-e-especular-warren-buffett-explica/. Acesso em: 16 nov. 2019.

No entanto, na definição de dicionário[4] são apresentados vários significados para a palavra "especular":

> **Especular**
> es·pe·cu·lar (2)
>
> (verbo transitivo direto e verbo intransitivo )
> 1. Estudar algo com atenção e minúcia, do ponto de vista teórico: *Os cientistas especularam as causas do fenômeno. Especulou muito e finalmente deve divulgar suas novas descobertas em biologia marinha.*
>
> (verbo intransitivo )
> 2. Dedicar-se a longas e profundas meditações que não tendem a produzir resultados práticos; cismar, cogitar, meditar: *Deixemos de especular vagamente, ajamos.*
>
> (verbo transitivo indireto)
> 3. Colher informações minuciosas acerca de alguma coisa: *Especulou sobre a vida anterior do seu pretendente.*
>
> (verbo transitivo indireto)
> 4. COM. ECON. Negociar no mercado de capitais ou câmbio com o objetivo de auferir lucros, aproveitando-se de uma situação temporária desse mesmo mercado.
>
> (verbo transitivo indireto)
> 5. ECON. Apostar na bolsa de valores ou de mercadorias.
>
> (verbo transitivo indireto)
> 6. Lançar mão de recursos especiais para iludir alguém em proveito próprio: *Especulou com a ingenuidade de agricultores pobres e desinformados.*
>
> (verbo transitivo direto e verbo intransitivo )
> 7. Conjeturar de modo indiscreto e malicioso, sem base em dados reais: *Todos especulam se o deputado está ou não envolvido em corrupção. Vivem a especular, mas não conseguem provar nada que o incrimine.*
>
> (verbo transitivo indireto)
> 8. Aproveitar-se de posição pessoal privilegiada para obter vantagens: *Especula com o fato de ser filho do governador para favorecer amigos e parentes.*

---

[4] MICHAELIS – **Dicionário Brasileiro da Língua Portuguesa**. Ed. Melhoramentos Ltda. 2019. Disponível em: https://michaelis.uol.com.br/palavra/8Y2X/especular-2/. Acesso em: 16 nov. 2019.

Etimologicamente,[5] a palavra "especular" vem do latim *speculari,* que significa "observar" ou "examinar". O sentido negativo do termo teria surgido a partir da conexão entre a ideia de "examinar/conjecturar" e a ideia de "elaborar hipóteses desprovidas de informações". A partir do desenvolvimento do mercado de ações, por sua vez, a palavra "especular" passaria, também, a referir a atitude de "tentar prever a evolução de uma empresa ou negócio". Há amparo, ainda, para a ligação etimológica com a palavra *specula* (lugar de observação), avançando para *speculationis* (espionagem), fator que transforma a simples contemplação em um ato nebuloso.[6] Ao cabo, todos esses significados se relacionam, com maior ou menor intensidade, carregando suas marcas culturais. O amálgama semântico daí derivado tem importância para a compreensão da natureza do *day trade* e de qualquer tipo de investimento.

Conforme lição de Assaf Neto (2018), *"o investimento é uma probabilidade de ganho"*, e é por isso que o mesmo autor sentenciou: *"tudo na vida é um investimento"*. De fato, seja quando alguém se dedica a um amor, seja quando adota posturas de prevenção à sua saúde, ou ainda quando opta por um produto financeiro para almejar lucro, o que se está fazendo são escolhas com o intuito de auferir resultado proveitoso no futuro – ou seja, um investimento.

Ocorre que o futuro é indefinido, de modo que as decisões, embora fundamentadas e racionalizadas, terão sempre uma parcela de aposta do investidor. Não há garantia plena e absoluta de retorno positivo, em qualquer investimento financeiro que seja. Nem mesmo um título público, emitido pelo governo da economia mais consolidada, é considerado integralmente seguro, já que *"não existe operação totalmente livre de risco"* (RASSIER, 2018). Essa componente de incerteza obriga os investidores à prévia ponderação, ao juízo crítico preliminar e, em boa dosagem, a um exercício controlado de futurologia, em que se avalia a probabilidade do que pode vir a acontecer para, enfim, tomar a decisão aparentemente mais favorável no presente. Por conta disso, se mostra acertado deduzir que a especulação é uma condição inerente a qualquer decisão de investimento, ao consubstanciar a dimensão conjectural do lucro futuro.

Como bem observado por Abreu (2018), *"os investimentos seguem traços culturais"*, o que é perceptível no âmbito das operações financeiras do mercado de capitais brasileiro. Em um país que ainda engatinha em termos de incursão em bolsa, são notáveis o incentivo e o endosso muito maiores à prática do chamado *buy and hold*, onde se compram ativos para mantê-los em carteira, do que às operações rápidas de *day trade*, que têm duração intradiária e podem ser realizadas, até mesmo, em poucos segundos. Ocorre que, embora o tempo para amadurecer o lucro (ou prejuízo) seja uma diferença impactante entre esses dois tipos de operação, no racional, ambas dependem de um mesmo princípio: estimar probabilidades de ganhos futuros.

---

[5] DICIONÁRIO ETIMOLÓGICO: **Etimologia e origem das palavras.** Ed. 7Graus. 2019. Disponível em: https://www.dicionarioetimologico.com.br/especular/. Acesso em: 16 nov. 2019.
[6] RODRIGUES, Sérgio. **O que a especulação tem a ver com o espelho.** Revista Veja. 2014. Matéria em *blog.* Disponível em: https://veja.abril.com.br/blog/sobre-palavras/o-que-a-especulacao-tem-a-ver-com-o-espelho/. Acesso em: 16 nov. 2019.

Se investidores de longo prazo acreditam em fundamentos de empresa e de macroeconomia como amparo da sua decisão, investidores de curto prazo acreditam em sinais gráficos ou seguem o fluxo de ordens para atuar. A alegação de aleatoriedade, ou de subjetividade na interpretação da movimentação de preços dos ativos, não está muito distante do perfil que conduz a principal metodologia de avaliação fundamentalista de empresas: o fluxo de caixa descontado. Por esse sistema de *valuation*, o potencial faturamento futuro, considerado até a perpetuidade, é trazido a valor presente com uma taxa de desconto apropriada, considerando aí embutida a estimativa de crescimento anual – tudo isso, tendo como baliza de comparação as expectativas macroeconômicas. A dificuldade de precisão do cálculo é bem ilustrada no ensinamento de Cavalcante (2018): *"O segredo do valuation está em sustentar as premissas com boas narrativas. Se o mercado acreditar nas narrativas da empresa, o seu valor se manterá alto"*. Na mesma linha, Miranda (2018) refere: *"Não há valor intrínseco, não há valor da empresa em si, há uma percepção sobre o valor da empresa. O valor é extrínseco, está na nossa cabeça, e não na empresa. Somos nós que atribuímos um cenário para a companhia"*.

Assim, tanto a compra de ativos para manutenção em carteira, como a compra e venda a curto prazo, até mesmo quando realizada em prazo intradiário, são todos investimentos igualmente baseados em especulação, costurados por argumentos e modulados pelo vetor tempo. A redução do tempo de investimento observada em operações curtas, entretanto, acarreta diferenças que impactam duramente na atividade do pequeno investidor.

No *day trade* há maior exposição às flutuações ocasionadas pelas chamadas operações ultrarrápidas. Tratam-se de negociações nascidas com o desenvolvimento de algoritmos poderosos, conhecidos como robôs HFT (negociação de alta frequência, na sigla em inglês). Tais mecanismos são utilizados por instituições financeiras e grandes *players* do mercado, realizando análises complexas e transações de grandes volumes em questão de segundos, o que gera mais liquidez no pregão e causa mudanças na volatilidade dos ativos/derivativos.

Nenhum ser humano tem paridade operacional com os robôs HFT, mas afastada a pretensão de concorrência direta, não se invalidam as chances de lucro dos pequenos investidores, já que a grande liquidez propicia entrada e saída de posições a qualquer momento do dia, e a alavancagem de capital, permitida no prazo intradiário, torna as operações atraentes em qualquer espectro de volatilidade. Além disso, o acirramento da competição institucional e os elevados custos da tecnologia HFT são indicativos de que a supremacia robótica pode vir a perder sustentabilidade em um futuro próximo (OSIPOVICH, 2017).

Outro ponto importante na comparação com os demais investimentos é que o *day trade* afeta, de forma mais dramática, atributos comportamentais e de controle emocional do investidor. Enquanto as operações mais longas podem ser planejadas, programadas, revisadas e executadas com maior liberdade temporal e estudos mais aprofundados, as operações curtas exigem atuação dinâmica e expressa, com tomadas de decisão imediatas, baseadas em análise de dados em tempo real. A pressão do tempo afeta o campo psicológico do investidor, não só pela necessidade de estar sempre pronto para decidir, mas também pelo dever de digerir e administrar de forma célere os resultados, sejam bons ou ruins. Os constantes e intensos estímulos emocionais funcionam como gatilhos para comportamentos incongruentes, pois incitam à prática de atos instintivos e sem respaldo estratégico. Nessa direção, Mosca (2018) deixa claro que *"toda a tendência comportamental que nós temos – inata, inconsciente, instintiva –, ela é ruim na hora que a gente vai gerir investimentos"*. Com a mesma autoridade, Amorim (2018) explica que *"muitas vezes, o que nos leva a fazer bobagem com o dinheiro são as nossas emoções"*.

Diante de tal panorama, é curioso que venha aumentando o número de interessados em fazer operações *day trade* no Brasil. Possivelmente, essa circunstância é resultado do crescente número de pessoas físicas que estão adentrando ao mercado de bolsa ao longo dos últimos anos.[7] Como consequência, por incidência direta da lei de oferta e procura, é natural que se multipliquem os cursos, palestras, entrevistas, material bibliográfico e, em geral, comércio de produtos e serviços voltados ao público entusiasta da prática de *day trade* – erroneamente divulgada e interpretada como sendo uma promessa de rápida escalada para a autossuficiência econômica e profissional. O saldo desse movimento é que existe, atualmente, um emaranhado de teses, conhecimentos e técnicas jogados na rede, imiscuídos e disseminados publicamente, acessíveis por qualquer indivíduo que utilize internet – o que, na maioria das vezes, se apresenta como um horizonte caótico e nocivo.

---

[7] LAPORTA, Taís. **Bolsa atinge a marca de 1,5 milhão de investidores**. Revista Exame. 2019. Matéria jornalística. Disponível em: https://exame.abril.com.br/mercados/bolsa-brasileira-atinge-a-marca-de-15-milhao-de-investidores/. Acesso em: 16 nov. 2019.

A proteção dos investidores no mercado de capitais é uma das atribuições da Comissão de Valores Mobiliários – CVM, o que ganha relevo em face da ainda incipiente educação financeira dos brasileiros em comparação a outras nações, como os EUA. Nessa linha, a pedido da CVM, recentemente foi publicado o estudo intitulado "É possível viver de *day-trading?*" (CHAGUE e GIOVANNETTI, 2019), o qual retrata, com base empírica, a dificuldade de pessoas físicas em obter lucro significativo e constante com a atividade de *day trade*. Foram apresentadas evidências de que tal prática traz perdas financeiras que se agravam com a repetição, não havendo sentido econômico na escolha dessa profissão.

O estudo é um dos poucos dedicados ao assunto no Brasil, e serve como alerta aos investidores em geral, por trazer à tona uma realidade sombria para a atividade de *day trade*, bastante distante do referencial de liberdade e independência financeira propalado na mídia. Porém, referido *paper* contém tópicos passíveis de problematização, de tal sorte que se vislumbram fundamentos para questionar e relativizar a universalização das suas conclusões.

### 3 – Delimitação do objeto de análise

Na seção anterior foram apresentadas proposições teóricas sobre as operações *day trade*, buscando ampliar o olhar sobre a natureza dessa atividade. Considerando a publicação do artigo "É possível viver de *day-trading?*" (CHAGUE e GIOVANNETTI, 2019), que se funda em bases empíricas, serão adiante abordados alguns dos seus elementos, em reflexão crítica.

O objeto a seguir analisado, portanto, se restringe à descrição daquela pesquisa e às suas conclusões. A base de dados amostrais que a subsidia não é aqui alvo de exame, pois cedida diretamente pela CVM, estando fora de cogitação qualquer dúvida palpável quanto à sua fidelidade descritiva.

O ponto é que, no campo metodológico da pesquisa, se verifica a eleição de premissas pouco justificadas, que encampam conceitos genéricos e suscetíveis a subjetivismos. Quanto à epistemologia, é questionável a validação de uma pesquisa dessa envergadura com base apenas em dados operacionais, sem uma abordagem complementar centrada nos indivíduos que compõem a população estudada. Ainda, do ponto de vista inferencial, as amostras de dados e os recortes de observação propostos se revelam muito restritos para a pretendida universalização das conclusões. Assim, com amparo na ideia de que *"a decisão em estatística é complexa porque está relacionada com todas as fases do processo de investigação e, principalmente, porque tem implicações no desenho de novos estudos"* (LOUREIRO e GAMEIRO, 2011, p. 160), a análise adiante imprimida tenciona discutir a pesquisa e colaborar para a evolução do tratamento científico dispensado ao tema.

Em cotejo analítico ao *paper*, serão apresentadas informações extraídas de outros artigos que circundam o mesmo assunto, além de menções a dados de pesquisas e relatórios do Instituto Brasileiro de Geografia e Estatística – IBGE, da Fundação Getúlio Vargas – FGV, e de outras fontes pontualmente referidas. Todo este referencial foi consultado diretamente em formato digital, estando disponível para acesso pela internet.

### 4 – Análise empírica

O exame de dados principia pela identificação do objetivo do *paper* em comento, definido pelos autores como sendo o de preencher uma *"lacuna informacional para auxiliar a tomada de decisão das pessoas que consideram a atividade de day-trading como uma alternativa profissional"* (p. 03). Complementarmente, ao longo do artigo se pode encontrar outra definição de objetivo: *"nosso objetivo é estudar a sobrevivência e performance dos day-traders que entram no mercado"* (p. 07). Ainda, um terceiro objetivo traçado é: *"O objetivo central deste trabalho é tentar responder à pergunta: é possível viver de day-trading?"* (p. 11).

De início, o significado das expressões "possível" e "viver de" não se mostra suficientemente delimitado. Em relação ao termo "possível", o resultado empírico daquele estudo demonstra que algumas pessoas – embora poucas – apresentam, sim, resultados lucrativos e contínuos fazendo *day trade*, de modo que a pergunta-título comportaria resposta positiva. Não é essa, todavia, a conclusão veiculada, indicando que a pergunta está menos voltada a uma averiguação de possibilidade, e mais direcionada a uma medição de probabilidade, ou dificuldade. Por outro lado, a expressão "viver de" também causa imprecisão, por ser deveras genérica e dar margem ao subjetivismo. No bojo daquele texto é utilizado o parâmetro monetário de lucro/sucesso como sendo a receita média de R$ 300,00 por dia, tida como *"renda significativa"* (p. 23), mas não é apresentado fundamento socioeconômico a ancorar tal cifra. Vale notar que o salário mínimo nacional é de R$ 998,00 no ano de 2019; a média da renda domiciliar *per capita* no Brasil, medida pelo IBGE para o ano de 2018, foi no patamar de R$ 1.373,00;[8] e a média geral de salários observada no país e medida pelo IBGE no ano de 2019 fica na faixa de R$ 2.270,00.[9] Essa questão ganha especial relevo em razão da segunda definição de objetivo mencionada naquele *paper*, em que se busca estudar a *"sobrevivência"* dos *day traders*. Pelas métricas estabelecidas no estudo, eventual *day trader* que tivesse lucrado, em média, R$ 299,00 por dia (multiplicando essa média por 21 dias úteis, chega-se a R$ 6.279,00 por mês), ficaria abaixo do piso estabelecido como exitoso – em que pese ostente renda considerável para sobreviver e seguir atuando na atividade diariamente. Ainda, em simples hipótese, um casal que atue exclusivamente operando *day trade*, onde cada um aufira, em média, R$ 191,00 por dia, alcançaria renda familiar de R$ 8.022,00 por mês – mas estaria muito longe da estatística de sucesso considerada no estudo. Essas incongruências sugerem que as balizas para responder à pergunta-título do *paper* se apresentam instáveis.

Seguindo o mote dos objetivos delineados naquela pesquisa, a medição da *"performance dos day-traders que entram no mercado"* não se mostra uma alternativa consistente para pontuar sobre a viabilidade da profissão, pois diz muito pouco sobre a condição de um *day trader* profissional e experiente, que "vive de" operações. Se há a pretensão de *"auxiliar a tomada de decisão"* daqueles que pretendem uma nova *"alternativa profissional"*, então é incoerente que esteja privilegiada, no bojo da estatística, a performance de amadores e novatos. Como consta descrito na pesquisa, foram desconsiderados da base de dados os *day traders* que tinham histórico de operações anteriores ao ano de 2013, fins de garantir que apenas quem recém iniciou sua atividade estivesse na amostra. Tal metodologia é deveras discrepante da utilizada no estudo *"Do Individual Day Traders Make Money? Evidence from Taiwan"* (BARBER, LEE, LIU e ODEAN, 2004), em que foi feito justamente o contrário: as medições excluíram *day traders* pouco ativos para garantir a medição de operadores com histórico de grande volume negocial. De fato, avaliar a possibilidade de se auferir renda suficiente para viver, em uma profissão específica, exige o estudo da capacidade de um profissional já em atividade plena, contabilizada em período de tempo mais significativo. Seria inconcebível a pretensão de avaliar se há sentido econômico na profissão, por exemplo, de um médico, usando como amostra da pesquisa a atividade e a remuneração de médicos-residentes ou estudantes de medicina. Coincidência ou não, o citado estudo sobre

---

[8] Disponível em: https://g1.globo.com/economia/noticia/2019/02/27/renda-domiciliar-per-capita-no-brasil-foi-de-r-1373-em-2018-mostra-ibge.ghtml. Acesso em: 16 nov. 2019.
[9] Disponível em: https://www.correiobraziliense.com.br/app/noticia/economia/2019/03/18/internas_economia, 743567/renda-de-servidores-eleva-a-media-salarial-do-pais.shtml. Acesso em: 16 nov. 2019.

*day traders* em Taiwan apresenta resultado estatístico de lucratividade 100% maior, alcançando o percentual de 18%-20% dos *day traders*, contra o percentual de 9%-12% encontrado no estudo brasileiro.

Os percentuais de *day traders* efetivamente lucrativos, apresentados em diversos estudos, são notavelmente baixos. Entretanto, é comum passar desapercebido que também é baixo o índice de êxito em outras profissões. Tome-se, a título de exemplo, a opção de empreender conduzindo pequenas empresas no Brasil. Conforme relatório sobre Demografia das Empresas e Estatísticas de Empreendedorismo, publicado pelo IBGE em 2019 e relativo ao ano de 2017,[10] se verifica uma taxa de sobrevivência empresarial decrescente ano a ano. Apenas 33% das empresas sem empregados chegam vivas ao quinto ano de atividade. E referido relatório não garante que estas empresas sobreviventes estejam gozando de plena saúde financeira e gerando lucros no quinto ano, já que não analisa a rentabilidade, mas sim o tempo de atividade. Nem por isso, todavia, se poderá concluir que a opção pelo empreendedorismo seja uma escolha sem sentido econômico. Em relação a profissionais liberais, se encontram cenários igualmente desafiadores. Exemplificativamente, a Fundação Getúlio Vargas – FGV divulgou relatório, no ano de 2016,[11] contendo a taxa de aprovação no Exame de Ordem da OAB ao longo de dezesseis edições da prova. Em proporção ao número de inscritos em cada exame, apenas 22,5% são aprovados. Vale destacar que todos os aprovados precisaram, previamente, cursar faculdade de Direito durante quatro a cinco anos, para só então estarem aptos a fazer o referido exame de ordem. Ao cabo, os 22,5% de aprovados estão apenas conseguindo se habilitar para iniciar a atuação profissional, não significando que imediatamente passarão a auferir renda desta profissão. Nem por isso, contudo, se poderá concluir que a opção pela carreira jurídica seja uma escolha sem sentido econômico.

Avançando desta linha, também merece ser problematizada, na pesquisa sob análise, a premissa relativa ao tempo necessário para a curva de aprendizagem, indicado como sendo "*de aproximadamente 1 ano*" (p. 03). Embora seja notória a existência de opiniões nesse sentido, disseminadas em portais de notícias e conteúdo financeiro, o fato é que não existe padronização, nem estatística, sobre o tempo que cada indivíduo necessita para o aprendizado. Em qualquer área do conhecimento que demande capacidade técnica, um bom profissional precisa de significativo tempo para se aprimorar, o qual vai variar de pessoa para pessoa, conforme grau de interesse, dedicação, capacidade, investimento. Para a compreensão de conhecimentos técnicos, o desenvolvimento de habilidades e o treino prático de estratégias, não seria extravagante imaginar uma curva de aprendizado similar à duração de qualquer curso em ensino superior, que leva em torno de quatro anos. Afinal, via de regra, só depois da construção de bases sólidas uma profissão pode ser exercida com excelência. Portanto, a inexistência de resultado econômico significativo após um, ou dois anos de atividade, pouco esclarece quanto à sustentabilidade financeira da atividade profissional de *day trader*. Não há motivos para imaginar que essa ocupação demande menos esforço do que o exigido de um empresário ou profissional liberal, no tocante ao tempo dedicado, ao planejamento, ao aprendizado, à maturação e, até mesmo, ao capital de investimento. Acreditar que seja possível estudar por um ano e cumprir uma "curva de aprendizagem" predefinida, passando a rentabilizar R$ 300,00 por dia sem antes investir/gastar dinheiro, soa como utópico e irreal.

---

[10] Disponível em: https://biblioteca.ibge.gov.br/visualizacao/livros/liv101671.pdf. Acesso em: 16 nov. 2019.
[11] Disponível em: https://fgvprojetos.fgv.br/sites/fgvprojetos.fgv.br/files/oab_3_edicao_v4_web_espelhado.pdf. Acesso em: 16 nov. 2019.

Ainda na parte metodológica do *paper*, se vislumbra baixa representatividade na amostra selecionada, a qual abrange apenas as operações *day trade* envolvendo derivativos específicos (mini-índice e mini-dólar), o que influencia sobremaneira os resultados da pesquisa. Primeiro, porque esses derivativos concentram a maior liquidez da bolsa, significando grande confluência de investidores e de estratégias em operação – o que transforma o ambiente negocial em verdadeira selva, desestabilizando os menos experientes e lhes gerando perdas. Segundo, porque referidos derivativos são índices futuros, característica que resulta em volatilidade alta e ocasiona movimentos mais dinâmicos e inconstantes do que no mercado acionário à vista – o que dificulta as análises e as estratégias de profissionais menos treinados, se traduzindo em alto índice de erros. E terceiro, porque embora os derivativos sejam alvo de predileção dos *day traders*, não é regra que o profissional atue estritamente nesse tipo de mercado, sendo coerente crer na exploração das oportunidades de forma geral na bolsa – inclusive, formando um sistema de *hedge* em que determinadas operações façam contrabalanço a outras. No caso dos dados empíricos apresentados no *paper*, a desconsideração de operações *day trade* feitas com outros ativos pode distorcer o resultado lucro, eis que *day traders* com prejuízo nos derivativos podem ter tido proveito econômico maior em ativos acionários, por exemplo. Na contramão dessa abordagem, outros estudos sobre o tema levaram em consideração todas as operações *day trade*, independentemente do ativo ou derivativo objeto, conforme se percebe em *"An Analysis of Public Day Trading at a Retail Day Trading Firm"* (JOHNSON, 1999), e em *"Do Individual Day Traders Make*

*Money? Evidence from Taiwan"* (BARBER, LEE, LIU e ODEAN, 2004). Em ambos os estudos o percentual de *day traders* lucrativos é superior ao verificado no estudo brasileiro.

A questão epistemológica, que diz com a validade ou possibilidade do conhecimento, é outro ponto sujeito a debate no âmbito da pesquisa sob apreço. Isso porque, não parece suficiente que apenas os dados operacionais de negociação forneçam satisfatoriamente uma resposta conclusiva à abrangente pergunta sobre a viabilidade econômica de escolher o *day trade* como profissão. Os dados analisados não qualificam os detalhes de cada investidor componente da amostra, para saber quem está "tentando viver de" *day trade,* não estando demonstrado o grau de empenho pessoal e dedicação empregados. Obviamente, eleger uma profissão não se resume a praticá-la de ofício. Acessar o *home broker* e operar derivativos é algo que pode ser feito repetidas vezes sem estudo, sem critérios, por insistência, por vício, dividindo a atenção com outra tarefa e sem foco determinado. Está sempre presente o interesse em auferir lucro, sem dúvida – mas nos cassinos também existe este interesse, só que é diferente de pretender tornar isso uma profissão. Os dados disponíveis não mostram, entretanto, quantas pessoas conduziram esse processo com seriedade, fazendo cursos, lendo, treinando e respeitando métodos. Seria epistemologicamente mais valioso – embora mais trabalhoso – se a pesquisa equacionasse os indivíduos da amostra pelo seu nível de comprometimento, medido em número médio de horas diárias consumidas, realização de cursos, regime de dedicação (se exclusiva, ou não), ou seja, alguns poucos dados concretos, passíveis de serem extraídos por entrevista ou pesquisa em formulário eletrônico, sistemática esta já utilizada com êxito no trabalho de Lukmann (2018). A ausência de elementos objetivos acaba forçando à adoção de premissas subjetivas para a pesquisa, a exemplo da seleção de indivíduos componentes da amostra, que compreende os que *"fizeram day-trades em mais de 300 pregões"* (p. 03), já que, segundo os autores, *"com alguma liberdade de interpretação, definimos essas pessoas como as que levaram à série o conselho de passar pelo suposto período de aprendizado e continuaram na atividade"* (pp. 03/04).

A propósito, merece um parágrafo de reflexão, também, a premissa que relaciona o número de pregões com o período de aprendizado do *day trader*. A pesquisa ora apreciada afirma que *"um total de 19.696 pessoas começaram a fazer day-trade em mini índice em 2013, 2014 e 2015. Dessas, 1.558 fizeram day-trades em mais de 300 pregões"* (p. 03). E, mais adiante naquele texto, é asseverado que *"das 19.696 pessoas que começaram a fazer day-trade em mini índice entre 2013 e 2015, 18.138 (92,1%) desistiram, umas mais cedo, outras mais tarde. Das 1.558 pessoas que passaram pelo período de aprendizado de 1 ano e* *continuaram a fazer day-trade, a grande maioria teve prejuízo"* (p. 04). Ou seja, segundo a pesquisa, quem atuou em menos de 300 pregões desistiu da atividade. Ocorre que, se não há dúvida de envolvimento com a atividade daqueles que atuaram em mais de 300 pregões (o que demonstra persistência por mais de 300 dias), o contrário não é necessariamente verdadeiro. Tanto para os aspirantes que estão entrando no mercado, como para os *day traders* profissionais que já atuam com consistência, não há evidências de que o êxito aumente em razão do aumento da frequência de exposição em bolsa. Pelo contrário: operar o dia inteiro, ou todos os dias, indiscriminadamente, acaba gerando mais taxas de corretagem e expondo o investidor ao risco em situações menos favoráveis. Um *day trader* experiente é aquele com alta capacidade para selecionar as operações de melhor risco-retorno, descartando as demais – e isso significa operar em menos pregões. Por outro lado, o já disseminado uso de programas simuladores de mercado, para teste em contas de demonstração, é um recurso tecnológico essencial, que permite dedicar tempo aos estudos, à montagem de estratégias e ao treino prático sem arriscar o capital do investidor na bolsa – e isso significa operar em menos pregões. Nesse sentido, um *day trader* (novato ou experiente), pode estar envolvido com o mercado, se dedicando com afinco à profissão, por dois ou três anos, inclusive obtendo lucros consideráveis, e ainda não ter atingido o número de 300 pregões reais com seu capital – estando, assim, excluído das estatísticas medidas no *paper* em comento.

No campo da inferência estatística, chama a atenção a construção do entendimento de que o *day trader* não evolui com o tempo e, em verdade, piora. Considerando a "Tabela 1" apresentada naquela pesquisa (p. 14), que retrata operações com o derivativo do mini-índice, vê-se que, após 250 pregões, os *day traders* mais lucrativos aumentaram seus lucros iniciais, embora essa população represente apenas 10% do total de *day traders*. Aqueles insertos no percentil 90% aumentaram em 1.950% a média de seus lucros diários (saindo de R$ 0,20 para R$ 3,90), e aqueles insertos no percentil 99% aumentaram em 25,89% a média de seus lucros diários (saindo de R$ 128,20 para R$ 161,40 diários). Os dados que amparam a conclusão veiculada no *paper*, entretanto, estão contidos na "Tabela 2" e na "Figura 5", que retratam, respectivamente, operações com o derivativo mini-dólar e nos derivativos agregados, onde houve efetiva piora no desempenho após 250 pregões. De todo modo, as variações de desempenho verificadas, sejam elas positivas ou negativas, não conduzem, por si só, a uma conclusão definitiva. Diversos estudos já foram realizados na tentativa de demonstrar o aprendizado de investidores com a própria experiência, havendo suporte para opiniões dissonantes, como se verifica em *"Do Individual Investors Learn from Their Trade*

*Experience?"* (NICOLOSI, PENG e ZHU, 2004), e também em *"Do Day Traders Rationally Learn About Their Ability?"* (BARBER, LEE, LIU, ODEAN e ZHANG, 2017).

Derradeiramente, o viés das conclusões apresentadas no *paper* também admite digressão. Embora a afirmativa de que *"não faz sentido, ao menos econômico, tentar viver de day-trading"* (p. 23) tenha amparo do significante estatístico, e ainda que se admita uma necessária margem de erro no levantamento, sua universalização sem ressalvas induz à inverídica ideia de que esteja comprovado que ninguém consegue viver adotando a profissão de *day trader* – o que é contrariado pelos próprios dados mensurados na pesquisa. De outro turno, a afirmativa de que *"à medida que o daytrader vai persistindo na atividade seu desempenho tende a ir piorando"* (p. 23) também se mostra questionável, eis que se baseia em duas das observações (mini-dólar e agregado dos derivativos), havendo perfil evolutivo na outra observação (mini-índice). E a base empírica para subsidiar a conclusão pela existência, ou não, de aprendizado, também está limitada a apenas estes dois derivativos específicos, passando ao largo de considerar a íntegra das operações de cada *day trader* na aferição de resultados. Embora a inferência estatística da pesquisa possa ser defendida como hígida, do ponto de vista estritamente formal, o modo como foram exprimidas as conclusões prejudica a compreensão dos resultados, ou de parte deles.

### 5 – Resultado da análise

A análise do artigo "É possível viver de *day-trading?*" (CHAGUE e GIOVANNETTI, 2019) demonstra que a conclusão nele veiculada – *"não faz sentido, ao menos econômico, tentar viver de day-trading"* – não pode ser tomada como sentença determinística ou consequência predefinida. A afirmativa tem lógica e se ampara em dados concretos, mas está construída a partir de frações da realidade, não contemplando a totalidade do fenômeno estudado – e, das partes de uma história, pode nascer outra história. Não há indicativo seguro de que a conclusão do *paper* se mostraria invariável ou exata caso fossem consideradas, coletadas e escrutinadas as informações sobre os demais elementos que moldam a experiência sensível de cada *day trader.*

Estudos estatísticos internacionais sobre *day trade* revelam dados que corroboram a necessidade de cautela com a rotulagem genérica dessa atividade enquanto economicamente deficitária, ou sem sentido para pessoas físicas. Ademais, a adoção do *day trade* como fonte de renda apresenta níveis de sobrevivência/mortalidade semelhantes a outras escolhas profissionais, com potencial de retorno financeiro capaz de suprir as necessidades de subsistência, em faixas de renda superiores à média salarial do país. Portanto, sendo o *day trade* uma oportunidade de investimento factível, com reais exemplos de consistência lucrativa, e havendo consenso de que a obtenção de sucesso nessa área é restrita e dificultosa, se mostra profícuo desvendar formas de otimização da atividade, de fomento das pesquisas científicas na área, e de incentivo à contínua formação do investidor interessado em aderir a essa prática.

### 6 – Especulações teóricas

Na seção anterior foram mencionadas pesquisas científicas com conclusões díspares acerca do aprendizado de um *day trader*. Dependendo do escopo de cada análise empírica levada a cabo, são revelados diferentes achados em relação à aprendizagem. Entretanto, existe um consenso, admitido até pelos mais céticos: *"The suggestion that investors learn from experience is neither novel nor controversial. Learning is a ubiquitous feature of human experience. From a welfare and policy perspective, the question is not whether investors learn, but how well they learn"* (BARBER, LEE, LIU, ODEAN e ZHANG, 2017, p. 03). Logo, há divergências quanto à qualidade, ou profundidade do aprendizado, notadamente considerando a maioria dos *day traders* – mas sem negá-lo, especialmente em relação aos *day traders* mais lucrativos.

Em linha antagônica, Chague e Giovannetti (2019) sustentam, no estudo examinado, que *"os resultados apresentados até aqui indicam que não deve haver aprendizado com o passar do tempo"* (p. 19). Na conclusão sintetizada no seu respectivo *abstract*, referem que *"o desempenho do day-trader não melhora à medida que ele persiste na atividade (na realidade, piora). Essa última evidência é crucial e vai de encontro à ideia, em geral propagada por especialistas das corretoras, de que day-traders melhorariam com a experiência e que, portanto, deveriam persistir"* (p. 01). Ou seja, a afirmativa é categórica no sentido de que *day traders* não aprendem ao persistirem, e que aumentam suas perdas ao longo do tempo.

Já que qualquer atividade racionalmente orientada pode ser alvo de aprendizado pelo ser humano, a negativa explicitada no *paper* em questão parece se dirigir mais à condição de irracionalidade do próprio mercado, cuja complexidade impediria o desenvolvimento de aptidões intelectuais a serem empregadas para a obtenção de lucratividade consistente. Assim considerada, a questão do aprendizado do *day trader* ganha ligeiro reposicionamento, se situando não na impossibilidade de adquirir conhecimento, mas na invalidade de um saber voltado ao fim de dominar um processo presumidamente estocástico. O aprendizado em *day trade* se assemelharia ao "aprendizado" em um jogo de cassino: onde a persistência, embora até possa vir a gerar lucro, não tem relação com a construção de conhecimento. Aliás, tal concepção foi exposta por um dos autores do *paper,* em matéria jornalística recentemente publicada.[12]

Um método para tentar esclarecer essa questão do domínio sobre a suposta aleatoriedade do mercado advém da especulação teórica. Tal proposta consiste na elaboração de hipóteses que confirmem ou rejeitem a possibilidade de contínua evolução dos *day traders*, à medida que sejam postas à prova pela racionalidade argumentativa (ALMEIDA, 2017). Não se tratando de uma investigação matemática, nem estatística, mas sim filosófica, importa aqui não almejar um resultado exato ou definitivo, mas que seja calcado na coerência. Para tanto, deve ser ponderado o vigor de cada fundamento, examinado o seu grau de proximidade com a realidade perceptível, e avaliado qual enunciado se revela mais factível e razoável de ser verificado como verdadeiro.

A primeira hipótese teórica, que confirmaria a possibilidade de aprendizado e de evolução ordenada da performance de um *day trader*, consiste em admitir que ele seja capaz de desenvolver ótima leitura dos movimentos do mercado intradiário, construir um exímio controle de riscos e, ao fim, lucrar de forma consistente. A segunda hipótese teórica, que rejeitaria a possibilidade de evolução ordenada do *day trader* ao longo do tempo, consiste em assumir que as negociações intradiárias têm natureza aleatória, com cenários imprevisíveis, impedindo o desenvolvimento de habilidades e estratégias para lucrar com consistência fazendo *day trade*. Uma terceira hipótese teórica, que também rejeitaria a possibilidade de evolução ordenada do desempenho do *day trader*, consiste em aceitar que ele seja capaz de construir conhecimento técnico e acumular experiência ao persistir na atividade, mas defender que seja inviável gerar êxito de forma constante com tais habilidades adquiridas, em razão da disputa direta e permanente com grandes *players* altamente qualificados.

---

[12] ELIAS, Juliana. **Day trade é cassino, muito mais sorte do que técnica, diz pesquisador.** Revista Exame. 2019. Publicação em periódico. Disponível em: https://exame.abril.com.br/mercados/day-trade-e-cassino-muito-mais-sorte-do-que-tecnica-diz-pesquisador/. Acesso em: 16 nov. 2019.

**6.1 – Teste da hipótese nº 1: "É possível desenvolver habilidades técnicas para lucrar consistentemente e evoluir de forma ordenada"**

Essa primeira hipótese teórica, que confirma a possibilidade de evolução performática ao longo do tempo, tem, a seu favor, um consenso científico: todos os estudos estatísticos até então desenvolvidos sobre *day traders* apresentam grupos de indivíduos, em percentuais maiores ou menores, que reportam lucratividade consistente e contínua.

Com isso, parte da hipótese nº 1, referente à existência do lucro consistente, está previamente validada por dados empíricos. Cabe, então, questionar se tal resultado deriva do desenvolvimento de habilidades cognitivas (havendo, assim, evolução ordenada), ou se existe outra causa fundante dos lucros observados. Para tanto, vão adiante delineados argumentos alternativos, com potencial para justificar os fatos, seguidos das respectivas objeções.

Opor que a consistência lucrativa de alguns *day traders* (pessoas físicas) possa se sustentar apenas pelo fator sorte, ou pelo acaso, é uma ideia que desafia a teoria das probabilidades, pois significaria aceitar que uma sequência de eventos aleatórios origine sempre um resultado certo e determinado – tal como ganhar na loteria todo mês. Por sua vez, opor que estes *day traders* são lucrativos porque manipulam o mercado a seu favor, seria um argumento ingênuo e rudimentar, considerando a dimensão dos valores que giram no pregão intradiário, e tendo em vista que somente grandes *players* institucionais e gestores de patrimônio detêm suficiente disponibilidade de capital para gerar movimentações relevantes nos preços, sendo desprezível a força de um pequeno investidor em relação a estas oscilações. Ainda, opor que a consistência operacional destes *day traders* decorra de informações privilegiadas, exigiria pressupor que cada indivíduo deste grupo possui uma rede maiúscula de informantes aptos a fornecer dados antecipados diariamente, sobre diversas posições de bolsa, o que, mesmo assim, só traria lucro se, após tornadas públicas tais informações, houvesse uma interpretação padronizada pelo restante dos *players*, que precisariam aderir sempre a favor do movimento previamente escolhido pelo *day trader*, lhe gerando, então, o efetivo proveito. Em outra frente, opor que este pequeno grupo lucrativo de *day traders* faça uso de alguma forma de trapaça operacional, ou aproveite alguma inconsistência técnica do sistema, significaria pôr em xeque a capacidade fiscalizatória de órgãos reguladores sólidos como a CVM, ou da própria B3 (atual nomenclatura da BOVESPA), cuja tecnologia e expertise são superiores àquelas detidas pelos investidores, sem contar a necessidade de driblar os sistemas das próprias corretoras que franqueiam o acesso do *day trader* à bolsa. Finalmente, opor que os *day traders* com consistência de lucro são pessoas físicas dotadas de talento inato para operar, além de ser um argumento bastante inverossímil, é também contraditório e, em última análise, confirmaria a hipótese nº 1, pois resultaria em admitir a existência de um método operacional (um racional técnico e comportamental) que torna a atividade lucrativa de forma constante.

Portanto, não se vislumbrando argumento suficientemente forte para a sua rejeição, a hipótese nº 1 se apresenta como verdadeira.

**6.2 – Teste da hipótese nº 2: "A natureza aleatória do mercado impede o desenvolvimento de habilidades e estratégias para lucrar com consistência"**

Essa hipótese nega a possibilidade de absorver experiências de mercado e construir conhecimento para aplicação futura, face à complexidade e inconstância dos movimentos intradiários dos preços. Assim sendo, os *day traders* pessoas físicas estariam em constante caça às bruxas, buscando compreender o incompreensível e queimando sua reserva de capital.

Os questionamentos oponíveis à hipótese nº 2 se relacionam, em parte, àqueles elencados para a hipótese nº 1. Afinal, a comprovação empírica de que existem, em maior ou menor número, *day traders* consistentemente lucrativos, força à procura de uma explicação para este dado. E, se for tomada como impossível a aquisição de habilidade a isso voltada, pouco resta como justificativa aos fatos, além das múltiplas oposições já acima rechaçadas.

Complementarmente, importa salientar que a hipótese nº 2 também peca porque a premissa de aleatoriedade do mercado (considerando que assim o seja) pode perfeitamente conviver com a premissa de evolução performática do *day trader*, uma vez que se incluem, dentre as habilidades operacionais a serem galgadas, o domínio comportamental e o refinamento da gestão de risco (WOLWACZ, 2010). O *day trader* consciente da complexidade do mercado, mesmo sem ostentar poder preditivo, exerce rígido controle de suas operações e, com isso, reduz perdas quando o movimento do mercado desborda da sua estimativa inicial e majora ganhos quando o movimento do mercado se consolida em favor da sua estimativa inicial. A detecção de momentos propícios e/ou confortáveis para abertura e fechamento de posições, além do autocontrole emocional para efetivamente executar o que se detecta, no momento correto – mesmo que isso signifique a decisão de não realizar nenhuma operação em dado momento – são instrumentos que permitem ao *day trader* acompanhar um processo supostamente estocástico (o movimento do mercado) e agir sobre ele de forma racional e determinada, imprimindo estratégias que gerem resultado favorável.

Por meio da analogia, com todas as ressalvas cabíveis, a situação pode ser comparada com a previsão do tempo, a condição do tráfego, ou a dinâmica do mar, eventos cujos dados exatos só são confirmados no momento em que ocorrem. Nunca há certeza prévia sobre a hora e o minuto em que vai chover; nem sobre a abrangência geográfica de um congestionamento que está se alastrando; nem sobre a velocidade e a força de uma onda em formação no mar. Entretanto, com conhecimento e experiência é possível ter estimativas e estar preparado para agir conforme o cenário se desvelar. Não fosse assim, seria necessário que todos carregassem diariamente um guarda-chuva; qualquer estagnação no trânsito sempre impediria a chegada ao destino; não haveria condições de praticar o *surf*. Nessa linha, é evidente que nenhum *day trader* consegue fazer apenas negócios exitosos, fundados em previsões exatas (aliás, estas são a minoria). Porém, é possível modificar a ação operacional de acordo com o cenário, fazendo com que a diferença entre o lucro e o prejuízo colhidos de múltiplas operações concretize um *spread* positivo, contínuo e consistente.

A hipótese nº 2 poderia ser verdadeira, por exemplo, no caso de ser falso o dado empírico, recorrente em diversas pesquisas, pelo qual sempre há algum grupo de *day traders*, maior ou menor, que apresenta lucro consistente. Nessa circunstância, seria plausível sustentar a ausência de evolução ordenada da performance. Contudo, crer que exista erro reiterado em diversos estudos científicos, sobretudo quando não há indícios que ponham em dúvida os dados de negociação neles coletados, não se revela uma alternativa fecunda.

O teste da hipótese nº 2 indica, pois, fragilidades, o que corrobora a confirmação da hipótese nº 1 como verdadeira.

## 6.3 – Teste da hipótese nº 3: "É inviável gerar êxito constante com as habilidades adquiridas, em razão da disputa direta com grandes *players* altamente qualificados"

Essa hipótese não contesta a possibilidade de aquisição de conhecimento e habilidades de negociação pelos *day traders* de varejo, mas se opõe à ideia de que tal bagagem possa fornecer consistência lucrativa de modo permanente, devido à alta competição com participantes ainda mais preparados, tais como instituições financeiras, gestores de capital, fundos, conglomerados, mesas proprietárias[13] e formadores de mercado.[14] Nesse contexto, os dados empíricos dos estudos estatísticos – que revelam grupos de *day traders* com consistência lucrativa – seriam fotografias limitadas, de um intervalo de tempo passageiro, que não se perpetua.

O raciocínio, apesar de convincente, é enganoso. Embora o mercado de bolsa seja plural e com franco acesso de qualquer parte do mundo, e apesar de haver gritante assimetria na qualificação dos diversos participantes e investidores, não há efetiva disputa entre um *day trader* pessoa física e os grandes *players*. A compreensão desse tema, contudo, exige que se faça um prévio nivelamento de informação quanto aos tipos de oferta e à forma de negociação em bolsa, conforme regras do respectivo manual operacional.[15]

Existem três maneiras de enviar uma intenção de negócio à bolsa: através de apregoação direta, de apregoação por leilão e de apregoação por oferta. Na apregoação direta, ocorrida apenas em situações especiais, a negociação já chega à bolsa com comprador e vendedor especificados pela corretora, envolvendo uma mesma quantidade de ativo/derivativo ao preço atual de mercado, o que configura um negócio fechado. Na apregoação por leilão, ocorrida quando não há fluxo normal de negociação, diversas intenções de compra e venda se acumulam e ficam com sua realização suspensa por um tempo determinado, após o qual são automaticamente executadas conforme o preço médio calculado, e no limite da liquidez existente para aquele nível de preço. Na apregoação por oferta, que é o fluxo normal de negociação, os participantes enviam à bolsa suas intenções de negócio, e a respectiva realização se submeterá à cronologia e aos níveis de preço constantes do livro de ofertas.

Considerando que o *day trader* depende da movimentação de preços para desenvolver sua atividade, os dois primeiros tipos de negociação acima referidos pouco lhe interferem. A apregoação direta, após validada pela bolsa, é automaticamente executada a preço de mercado, sem sequer ser registrada no livro de ofertas, ocasionando ínfima ou nenhuma movimentação de preço. Igualmente, a apregoação por leilão, inclusive por definição conceitual, não gera movimento no preço (os negócios ficam suspensos), o que só é retomado após o transcurso do prazo regulamentar e o restabelecimento do fluxo de negociação.[16] Assim, o que preponderantemente impacta a atividade do *day trader* é a execução de negócios referenciada pelo livro de ofertas – sendo pertinente explicitar a sua metodologia.

---

[13] Mesa proprietária *(Prop Trading)* é uma espécie de "empresa de traders, que atua no mercado financeiro, assim como qualquer trader faria no varejo convencional. A única diferença entre eles é que esses traders não operam com seus próprios dinheiros, mas sim com o da empresa ou Mesa Proprietária. Sendo assim, a mesa fornece a educação financeira, o capital e a infraestrutura e o trader opera e faz dinheiro de volta para ela" (SANTOS, 2019).

[14] Formador de mercado *(Market Maker)* é um agente que funciona "quase como um 'funcionário' da Bolsa. Ele é contratado para colocar ordens no mercado nos dois lados, tanto comprador quanto vendedor, e fica ali esperando contraparte" (FERREIRA, 2018?).

[15] Disponível em: http://www.b3.com.br/data/files/93/D2/40/3B/8AFE961023208E96AC094EA8/Manual% 20de%20procedimentos%20operacionais%20de%20negocia%C3%A7%C3%A3o%20da%20B3%20-%20Vers %C3%A3o%2008042019.pdf.pdf. Acesso em: 16 nov. 2019.

Cada interessado envia à bolsa sua intenção de negócio (compra ou venda) na forma de ordens, seja manualmente ou através de acionamento automático – *stop*. Se a ordem disparada ofertar vantagem ou, ao menos, equivalência de preço em relação à melhor ordem de contraparte já registrada no livro, a negociação é executada imediatamente. Do contrário, não ocorre negociação e a ordem disparada ficará, ela própria, registrada no livro para aguardar execução caso a cotação chegue ao encontro do preço ofertado. Os registros do livro são divididos por lado (compra ou venda) distribuídos por níveis de preço e organizados por sequência cronológica, com precisão de milissegundos, garantindo assim: i) preferência de execução às ordens que ofereçam melhor preço à contraparte; ii) preferência de execução às ordens registradas com maior antecedência em cada nível de preço. Ao cabo, a cotação só se movimenta conforme vai sendo exaurida a liquidez nos melhores níveis de preço.

Nivelado o conhecimento técnico, é de fácil percepção que o poderio econômico e a extrema qualificação de grandes *players* não lhes garante preferência formal na execução das suas ordens. Todas as ofertas apregoadas obedecem rigorosamente aos critérios de anterioridade e proveito acima destacados, sendo impessoal a escolha de quem converterá sua intenção em negócio. Desse modo, o pequeno investidor pode tranquilamente executar sua ordem com prioridade em relação à ordem de um grande banco, ou de uma competente mesa proprietária, pura e simplesmente por ofertar um preço mais favorável em relação à melhor ordem de contraparte, ou por já ter registrado sua ordem no livro com maior antecedência em determinado nível de preço. Tal sistemática inibe qualquer distinção no tratamento das ordens dos múltiplos agentes de mercado.

Paradoxalmente, a paridade quanto à execução de ordens reforça a ideia de disputa entre os participantes, caracterizada pela livre concorrência dos interessados em consumir liquidez de uma mesma faixa de preço. De fato, considerando que todo negócio é único (não pode ser realizado duas vezes), e considerando que as ofertas do livro não são ilimitadas, é coerente imaginar que haja disputa por oportunidade entre os participantes. Disso sobreviria a vantagem competitiva dos grandes *players,* sabidamente detentores de maior estrutura e qualificação para analisar, decidir e implementar o disparo de suas ordens à bolsa. Entretanto, a realidade prática tem suas sutilezas. Para aclarar a questão, é oportuno utilizar um exemplo caricato, extraído do mundo animal: é como acreditar que, em um mesmo ambiente, haja disputa direta por alimento entre um elefante e uma formiga, já que ambos comem folhas, e ninguém pode comer a mesma folha, e as folhas não são infinitas. Ocorre que, por óbvio, o elefante não disputa o seu alimento com a formiga; a alimentação de um não inviabiliza a do outro. É certo que a formiga precisa estar atenta aos movimentos do elefante, para não ser esmagada; deve conhecer as variáveis do ambiente, para saber se posicionar; mas ela não é o alvo do elefante e, agindo adequadamente, pode se aproveitar de folhas não consumidas por ele.

---

[16] No caso dos leilões, ressalve-se que pode haver reflexo na atividade do *day trader* em situações específicas, estritamente quando já há posição aberta ao tempo da instalação do leilão, ocasião em que haverá exposição ao *spread* entre a cotação imediatamente anterior ao leilão e a cotação de reabertura do fluxo de negociação.

No caso do mercado, essa simbiose é explícita ao se verificar o volume financeiro médio da negociação em bolsa, que atualmente beira os 20 bilhões de reais por dia.[17] Tamanha cifra atesta o imenso capital envolvido em operações pela coletividade de participantes. Acontece que o volume de negociação em cada ativo/derivativo não se fixa em um único e restrito nível de preço, pois se distribui conforme o interesse dos participantes ao longo do pregão. A magnitude desses movimentos faz com que os diversos níveis de preço percorridos apresentem farta liquidez em relação às diminutas posições detidas por pessoas físicas (da mesma forma que um único galho de árvore tem capacidade para alimentar centenas de formigas). Nesse sentido, pequenos investidores podem abrir ou fechar suas operações a qualquer instante, sempre encontrando contraparte disponível para fazer negócio pela cotação atual de mercado.[18] O preço de fechamento do negócio pelo pequeno investidor, embora possa sofrer pequenas variações conforme haja grande confluência de ordens em determinados níveis de preço, via de regra fica muito próximo à cotação em que houve o disparo da sua ordem. Por isso, independentemente da atuação simultânea de grandes *players,* o *day trader* de varejo pode decidir e agir negocialmente no instante em que desejar, normalmente encontrando contraparte a uma cotação equivalente, ou circunvizinha ao preço do último negócio realizado.[19]

---

[17] "O volume financeiro médio diário negociado na B3 no segmento de ações, que contempla o mercado à vista e derivativos sobre ações, subiu 84,5% em agosto ante o mesmo mês de 2018, para R$ 19,738 bilhões", conforme veiculado por Campos (2019), no portal Valor Econômico.

[18] Contribui para isso a implantação de uma nova modalidade de ordens, em fase de testes desde 05/08/2019, denominada RLP *(Retail Liquidty Provider),* em que é permitido à própria corretora atuar como contraparte dos seus clientes e lhes gerar vantagem quanto ao preço de negociação, lançando ordens que não são registradas no livro de ofertas, o que aumenta de forma significativa a liquidez existente em cada nível de preço, conforme especificações constantes do *website* da B3. (Disponível em: http://www.b3.com.br/pt_br/solucoes/ plataformas/puma-trading-system/para-participantes-e-traders/regras-e-parametros-de-negociacao/novo-tipo-de-oferta-retail-liquidity-provider-rlp/).

Para além de tudo o que já foi mencionado, também descabe falar em concorrência que afete a performance do *day trader* porque, em verdade, o seu êxito ou fracasso tem pouca relação com a execução, propriamente dita, das suas ordens. O que tem relevância no resultado da operação é a sucessão de eventos que ocorre "entre" as suas ordens (quando a posição já foi aberta e oscila pelo movimento da cotação). Ou seja, interfere pouco no *day trade* o procedimento mecânico de execução das ordens, que se deflagra na fração de segundo entre o disparo e a sua realização na bolsa; o mérito da operação depende muito mais do processo que se desenvolve depois de iniciada a operação e antes do disparo da ordem de fechamento. Por isso, a performance do *day trader* não sofre prejuízo em razão da quantidade, ou da qualidade dos participantes do pregão, tendo mais relação com a direção e a extensão do movimento do preço enquanto a operação está aberta – dinâmica que pode variar independentemente de quem esteja interagindo no mercado.

O raciocínio pode ser exemplificado através do conhecido jogo de azar dos cassinos, denominado Roleta Europeia, cujas regras e probabilidades matemáticas foram bem retratadas por Andrade (2017). A aposta plena (assim chamada quando feita em um único número) oferece a chance de êxito de aproximadamente 1/37, pois há 37 possibilidades de número alvo, mas apenas em uma delas haverá êxito do apostador. Se este raciocínio fosse aplicado ao mercado de bolsa, em função do número total de participantes globais, a chance de êxito de cada *day trader* seria absurdamente desfavorável, tornando fantasiosa a ideia de sucesso contínuo. Porém, o sucesso ou fracasso de uma operação de *day trade* depende basicamente do deslocamento direcional da cotação após a abertura da posição de compra ou venda – e não existem tantas alternativas de direção. O movimento do preço é binário, havendo só dois eventos possíveis: subir ou descer.[20]

---

[19] A viabilidade de, a qualquer momento do pregão, negociar em um nível de preço adjacente ao do último negócio realizado é aqui sustentada com foco no *day trader* pessoa física, e sempre em relação aos principais ativos/derivativos da bolsa, que apresentam grande volume de negociação e são dotados de altíssima liquidez. Já *players* institucionais ou gestores de vultoso capital, por precisarem atender a demandas em volumes mais significativos, bem como quaisquer investidores que negociem em ativos/derivativos de baixa liquidez, estarão propensos a se deparar com menor disponibilidade de ofertas em cada nível de preço, ficando expostos a maior risco de *spread* da cotação na execução de suas ordens.

[20] O movimento do preço é binário porque a inexistência de deslocamento da cotação, com manutenção do preço estagnado até o encerramento do *trade*, não faz movimento direcional, ficando assim excluída do raciocínio. De qualquer forma, mesmo se fosse considerada como um terceiro evento possível, teria pouca influência na análise teórica aqui desenvolvida. Isso porque, embora seja desfavorável para o *day trader* a sua ocorrência, por envolver custos operacionais, deve-se ter em conta que é rara a sua concretização (o preço normalmente oscila

A mesma condição binária também ocorre em uma aposta específica do jogo da Roleta: a escolha entre vermelho/preto. Neste caso, assumindo a aleatoriedade do jogo, a chance de êxito pode ser entendida como de 1/2, havendo probabilidade matemática de aproximadamente 50% para um resultado favorável à escolha prévia.[21] Tal panorama, se confrontado com a realidade bidirecional do preço de ativos/derivativos no mercado, gera uma incógnita: por que o *day trader* não encontra sucesso à proporção de 50%, e a consistência lucrativa aparece nas pesquisas em índices tão baixos?

Há diferenças importantes entre as duas situações. Primeiro que, no jogo da Roleta, não existe gradação do resultado, pois sempre é sorteado um número exato, equidistante do próximo, que é excludente de qualquer outro. Já na bolsa, os movimentos de preço não obedecem a equidistâncias, nem a prazos, havendo transições na intensidade com que a operação evolui em direção ao ponto de ganho ou de perda preestabelecido, sendo possível o encerramento do *trade* a qualquer momento pelo investidor. A segunda diferença é que, no jogo da Roleta, a aposta do vermelho/preto tem chance constante em 1/2, pois a regra de apostas não permite alterar isso. Já na bolsa, dependendo da diferença estipulada entre ganho/perda, o *day trader* pode alterar a razão de risco/retorno, estabelecendo, por exemplo, alvos de ganho maiores do que o ponto de saída por prejuízo, ou pontos de saída por prejuízo maiores do que os alvos de ganho – e, até mesmo, em casos mais graves, sequer estabelecer um ponto de saída por ganho ou prejuízo. A terceira importante diferença é que, no jogo da Roleta, não se pode intervir ou modificar as condições da aposta durante o processo, então o resultado observará estritamente o caráter aleatório do sorteio. Já no *day trade,* entre a abertura e o fechamento de uma posição podem ocorrer muitos movimentos e oscilações da cotação, com alternância de subida e descida, ou sequências unidirecionais, o que causa impacto emocional no investidor e o induz a interferir no processo, dando ensejo a: erros operacionais, interpretações ambíguas, precipitação ou demora na decisão, precipitação ou demora na implementação da decisão, negligência na gestão de risco, alavancagens e realizações parciais indevidas, má mensuração dos alvos, etc.. Em teoria, se fosse reduzida à neutralidade a interferência do *day trader*, com simples definição de pontos de ganho e perda equidistantes, deveria ser alcançada, na média e após uma série de repetições, uma

---

ao longo do pregão), além de haver diluição do custo de suas parcas incidências frente a todas as outras operações.

[21] A rigor, também pode sobrevir um terceiro resultado na aposta do vermelho/preto: o sorteio do número zero (cor verde), que favorece a banca. A chance de sua ocorrência é de 1/37, de modo que preponderam fortemente os resultados de vermelho ou preto. De qualquer forma, a probabilidade matemática exata de resultado na aposta do vermelho/preto é de 48,64% (ANDRADE, 2017).

probabilidade de sucesso em torno de 50% – supondo o mercado como aleatório. Isso demonstra que o alto nível de insucesso verificado nas pesquisas deriva diretamente da atuação e das escolhas do *day trader,* seja pela constante interferência em um processo supostamente aleatório, seja pela incompreensão de um processo que, talvez, não se mostre tão aleatório assim.

O fato é que, uma vez aberta a posição de *day trade*, só há dois eventos possíveis, em termos de movimento direcional de preço: subir ou descer. E sempre, durante o pregão, o *day trader* está constantemente exercendo a escolha binária de encerrar o *trade* ou seguir com a operação aberta. Este par de alternativas, que é o fundamento do fracasso/sucesso operacional, é imune à intervenção dos demais participantes do mercado, sejam eles grandes *players* qualificados, ou não. A disputa do *day trader* se dá, em verdade, contra as próprias convicções.

Portanto, a hipótese nº 3 não se demonstra adequada, o que reitera a confirmação da hipótese nº 1 como verdadeira.

**6.4 – Síntese da especulação teórica**

Os argumentos concretos que foram tecidos para verificar a validade de cada hipótese abstrata conduzem à identificação daquela que parece ser a mais aceitável em relação ao aprendizado do *day trader* pessoa física, qual seja, a hipótese nº 1: "é possível desenvolver habilidades técnicas para lucrar consistentemente e evoluir de forma ordenada".

Essa assertiva não significa dizer que a maioria dos *day traders* de varejo consiga evoluir, nem significa que seja simples e rápido construir desempenho sólido e constante. Indubitavelmente, a jornada se revela complexa e exige sacrifícios. Os números apresentados nos diversos estudos estatísticos já publicados sobre o tema servem como notória demonstração de que o *status* de sucesso só é alcançado por uma pequena fração de investidores, deixando um rastro de destruição quanto aos outros.

No entanto, constatar a inexistência de obstáculos intransponíveis para essa conquista representa subir um degrau crucial na perspectiva de quem pretende se profissionalizar na atividade de *day trader.* Ao menos do ponto de vista objetivo, a possibilidade existe. E é a competência subjetiva de cada investidor que definirá o grau de transmutação dessa realidade abstrata em realidade concreta.

**7 – Conclusão**

O título escolhido para o presente trabalho foi *"DAY TRADE* – Do outro lado das estatísticas", no intuito de evidenciar que as referidas operações financeiras contam com dimensões complementares à realidade empírica largamente difundida. Afinal, na outra ponta da base de dados estão as pessoas, em toda a sua complexidade e humanidade. Os indivíduos, com sua bagagem, suas vicissitudes e virtudes, são os responsáveis por endereçar os pontos da curva estatística, através do resultado de operações financeiras bem ou mal escolhidas – bem ou mal especuladas.

Sustentar que pessoas físicas estejam predestinadas ao déficit financeiro em operações *day trade*, e que devam se abster de enveredar por este caminho porque a repetição agravaria o seu prejuízo, não se afigura uma posição passível de validação por amparo exclusivo dos dados negociais de operações selecionadas. Tais medições ignoram o contexto, as metas, a experiência, o ânimo, a obstinação, a entrega, a resiliência, o talento – em última análise, a paixão com que se dedica à atividade cada potencial operador do mercado.

Em leitura alternativa, o que bem demonstram as estatísticas é a existência de pessoas, em um pequeno grupo, que, de forma inexplicada, fogem da curva e fazem dinheiro de forma constante. Seria muito proveitoso que o ímpeto de estudiosos do tema se voltasse a dissecar a atuação desses expoentes, para assim contribuir com o desenvolvimento dos demais integrantes da base de dados das pesquisas, ampliando a coletividade de pequenos investidores beneficiados economicamente com o *day trade*. Afinal, ao menos em teoria, é plausível admitir a possibilidade de efetiva evolução da performance operacional através da persistência na atividade.

É difícil encontrar respostas absolutas quando estão na mira do pesquisador questões objetivas e subjetivas, intrínsecas do ser humano. É salutar, entretanto, reconhecer e dar tratamento a tudo aquilo que seja capaz de interferir nos resultados da averiguação. Nesse sentido, é inegável que a brutalidade do mercado financeiro intradiário perpassa os limites do seu ambiente de bolsa: não raro, por sua complexidade fenomenológica, é capaz de criar obstáculos ideológicos e psicológicos mais prejudiciais ao sucesso do que as próprias barreiras técnicas da negociação.

# REFERÊNCIAS

ABREU, Edgar. **Produtos e Serviços Financeiros**. 2018. Aula do curso de pós-graduação em Finanças, Investimentos e Banking, da Escola de Negócios da PUC/RS.

AFFONSO, Rafael Bechara. **O impacto do High-Frequency Trading na relação volatilidade-preço no mercado acionário.** 2018. Monografia (PUC/MG). Disponível em: https://www.webartigos.com/storage/app/uploads/public/5c0/67d/3a1/5c067d3a1e521694246 337.pdf. Acesso em: 16 nov. 2019.

ALMEIDA, Aires. **Racionalidade argumentativa da Filosofia e a dimensão discursiva do trabalho filosófico**. 2017. Artigo científico. Disponível em: http://apfilosofia.org/wp-content/uploads/2017/09/AE-LÓGICA.pdf. Acesso em: 16 nov. 2019.

AMORIM, Ricardo. **Finanças Comportamentais**. 2018. Aula do curso de pós-graduação em Finanças, Investimentos e Banking, da Escola de Negócios da PUC/RS.

ANDRADE, Rafael Thé Bonifácio de. **A probabilidade aplicada aos jogos de azar**. 2017. Dissertação de mestrado (UFPB). Disponível em: https://repositorio.ufpb.br/jspui/bitstream/ tede/9474/2/arquivototal.pdf. Acesso em: 18 nov. 2019.

ASSAF NETO, Alexandre. **Investimentos I: teoria**. 2018. Aula do curso de pós-graduação em Finanças, Investimentos e Banking, da Escola de Negócios da PUC/RS.

BARBER, B. M.; LEE, Y-T.; LIU, Y-J.; ODEAN, T. *Do Individual Day Traders Make Money? Evidence from Taiwan*. 2004. Artigo científico. Disponível em: https://pdfs.semanticscholar.org/79ca/a03f933fd6d0c945fdfc9d8ae97b1b4fc2fb.pdf. Acesso em: 16 nov. 2019.

BARBER, B. M.; LEE, Y-T.; LIU, Y-J.; ODEAN, T.; ZHANG, K. *Do Day Traders Rationally Learn About Their Ability?* 2017. Artigo científico. Disponível em: https://faculty.haas.berkeley.edu/odean/papers/Day%20Traders/Day%20Trading%20and%20 Learning%20110217.pdf. Acesso em: 16 nov. 2019.

BRASIL. Lei nº 6.385, de 7 de dezembro de 1976. **Dispõe sobre o mercado de valores mobiliários e cria a Comissão de Valores Mobiliários.** Disponível em: http://www.planalto.gov.br/ccivil_03/leis/L6385compilada.htm. Acesso em: 16 nov. 2019.

B3 S/A – BRASIL, BOLSA, BALCÃO. **Manual de procedimentos operacionais de negociação da B3**. Versão 08/04/2019. Disponível em: http://www.b3.com.br/data/files/93/ D2/40/3B/8AFE961023208E96AC094EA8/Manual%20de%20procedimentos%20operacionai s%20de%20negocia%C3%A7%C3%A3o%20da%20B3%20-%20Vers%C3%A3o%200804 2019.pdf.pdf. Acesso em: 16 nov. 2019.

B3 S/A – BRASIL, BOLSA, BALCÃO. **Retail Liquidity Provider (RLP)**. 2019. Instruções em *website*. Disponível em: http://www.b3.com.br/pt_br/solucoes/plataformas/puma-trading-system/para-participantes-e-traders/regras-e-parametros-de-negociacao/novo-tipo-de-oferta-retail-liquidity-provider-rlp/.                    Acesso              em:            16            nov.            2019.

CAMPOS, Álvaro. **Volume médio diário no segmento de ações na B3 tem alta anual de 84,5%**. Valor Econômico. 2019. Matéria em periódico. Disponível em: https://valor.globo.com/financas/noticia/2019/09/11/volume-medio-diario-no-segmento-de-acoes-na-b3-tem-alta-anual-de-845.ghtml. Acesso em: 16 nov. 2019.

CAVALCANTE, Francisco. **Valuation: avaliação de empresas**. 2018. Aula do curso de pós-graduação em Finanças, Investimentos e Banking, da Escola de Negócios da PUC/RS.

CHACRA, Gustavo. **Operações em alta velocidade obtêm mais ganhos na Bolsa de Nova York**. Estadão. 2009. Publicação em periódico. Disponível em: https://economia.estadao.com.br/noticias/geral,operacoes-em-alta-velocidade-obtem-mais-ganhos-na-bolsa-de-nova-york,412118. Acesso em: 16 nov. 2019.

CHAGUE, F.; GIOVANNETTI, B. **É possível viver de *day-trading?*** 2019. Artigo científico. Disponível em: https://cointimes.com.br/wp-content/uploads/2019/03/Viver-de-day-trading-1.pdf. Acesso em: 16 nov. 2019.

D'ÁVILA, Mariana. **Você sabe a diferença entre investir e especular? Warren Buffett explica**. InfoMoney. 2016. Publicação em *site* de conteúdo e notícias. Disponível em: https://www.infomoney.com.br/onde-investir/voce-sabe-a-diferenca-entre-investir-e-especular-warren-buffett-explica/. Acesso em: 16 nov. 2019.

DICIONÁRIO ETIMOLÓGICO: **Etimologia e origem das palavras**. Ed. 7Graus. 2019. Disponível em: https://www.dicionarioetimologico.com.br/especular/. Acesso em: 16 nov. 2019.

ELIAS, Juliana. **Day trade é cassino, muito mais sorte do que técnica, diz pesquisador**. Revista Exame. 2019. Publicação em periódico. Disponível em: https://exame.abril.com.br/mercados/day-trade-e-cassino-muito-mais-sorte-do-que-tecnica-diz-pesquisador/. Acesso em: 16 nov. 2019.

FERREIRA, Janderson de Farias. **Como atuam os Market Makers?** L&S Educação. [2018?] Artigo em portal de conteúdo. Disponível em: https://ls.com.vc/educacao/artigo/como-atuam-os-market-makers. Acesso em: 16 nov. 2019.

FUNDAÇÃO GETÚLIO VARGAS – FGV. **Exame de Ordem em números**. 2016. Disponível em: https://fgvprojetos.fgv.br/sites/fgvprojetos.fgv.br/files/oab_3_edicao_v4_web_espelhado.pdf. Acesso em: 16 nov. 2019.

INSTITUTO BRASILEIRO DE GEOGRAFIA E ESTATÍSTICA – IBGE. **Demografia das empresas e estatísticas de empreendedorismo - 2017**. 2019. Disponível em: https://biblioteca.ibge.gov.br/visualizacao/livros/liv101671.pdf. Acesso em: 16 nov. 2019.

JOHNSON, Ronald L. *An Analysis of Public Day Trading at a Retail Day Trading Firm*. 1999. Artigo científico. Disponível em: https://docplayer.net/7803029-An-analysis-of-public-day-trading-at-a-retail-day-trading-firm.html. Acesso em: 16 nov. 2019.

KARNAL, Leandro. **Cultura, Política e Poder em Finanças**. 2018. Aula do curso de pós-graduação em Finanças, Investimentos e Banking, da Escola de Negócios da PUC/RS.

LAPORTA, Taís. **Bolsa atinge a marca de 1,5 milhão de investidores**. Revista Exame. 2019. Matéria jornalística. Disponível em: https://exame.abril.com.br/mercados/bolsa-brasileira-atinge-a-marca-de-15-milhao-de-investidores/. Acesso em: 16 nov. 2019.

LAPORTA, Taís. **Renda domiciliar per capita no Brasil foi de R$ 1.373 em 2018, mostra IBGE.** G1. 2019. Matéria jornalística. Disponível em: https://g1.globo.com/economia/noticia/2019/02/27/renda-domiciliar-per-capita-no-brasil-foi-de-r-1373-em-2018-mostra-ibge.ghtml. Acesso em: 16 nov. 2019.

LOUREIRO, L.M.J.; GAMEIRO, M. G. H. **Interpretação crítica dos resultados estatísticos: para lá da significância estatística**. Revista de Enfermagem Referência III nº 3. 2011. Artigo Científico. Disponível em: http://www.scielo.mec.pt/pdf/ref/vserIIIn3/serIIIn3a16.pdf. Acesso em: 16 nov. 2019.

LUKMANN, Pablo Muriel. **Gerenciamento de risco em operações de *day trade***. 2018. Trabalho de Conclusão de Curso (FGV-IDE). Disponível em: http://www.fgv.br/network/tcchandler.axd?TCCID=8159. Acesso em: 16 nov. 2019.

MASSARO, André. **Quer ser um day trader? Então faça o teste**. Revista Exame. 2012. Publicação em periódico. Disponível em: https://exame.abril.com.br/blog/voce-e-o-dinheiro/quer-ser-um-day-trader-entao-faca-o-teste/. Acesso em: 16 nov. 2019.

MAZZONI, Conrado. **Alta frequência deve girar 20% da bolsa até o fim do ano**. Valor Econômico. 2012. Publicação em periódico. Disponível em: http://www.valor.com.br/impresso/financas/alta-frequencia-deve-girar-20-da-bolsa-ate-o-fim-do-ano. Acesso em: 16 nov. 2019.

MICHAELIS – **Dicionário Brasileiro da Língua Portuguesa**. Ed. Melhoramentos Ltda. 2019. Disponível em: https://michaelis.uol.com.br/palavra/8Y2X/especular-2/. Acesso em: 16 nov. 2019.

MIRANDA, Felipe. **Investimentos**. 2018. Palestra ministrada em caráter de introdução ao curso de pós-graduação em Finanças, Investimentos e Banking, da Escola de Negócios da PUC/RS.

MOSCA, Aquiles. **Gestão de Recursos de Terceiros**. 2018. Aula do curso de pós-graduação em Finanças, Investimentos e Banking, da Escola de Negócios da PUC/RS.

NICOLOSI, G.; PENG, L.; ZHU, N. *Do Individual Investors Learn from Their Trade Experience?* 2004. Artigo científico. Disponível em: https://www.researchgate.net/publication/222836913_Do_Individual_Investors_Learn_from_Their_Trade_Experience. Acesso em: 16 nov. 2019.

OSIPOVICH, Alexander. **Transação de alta frequência sofre para manter ganhos**. Valor Econômico. 2017. Publicação em periódico. Disponível em: https://valor.globo.com/financas/ noticia/2017/03/22/transacao-de-alta-frequencia-sofre-para-manter-ganhos.ghtml. Acesso em: 16 nov. 2019.

PONDÉ, Luiz Felipe. **Filosofia do Dinheiro**. 2018. Aula do curso de pós-graduação em Finanças, Investimentos e Banking, da Escola de Negócios da PUC/RS.

RASSIER, Leandro. **Produtos e Serviços Financeiros**. 2018. Aula do curso de pós-graduação em Finanças, Investimentos e Banking, da Escola de Negócios da PUC/RS.

RODRIGUES, Sérgio. **O que a especulação tem a ver com o espelho**. Revista Veja. 2014. Matéria em *blog*. Disponível em: https://veja.abril.com.br/blog/sobre-palavras/o-que-a-especulacao-tem-a-ver-com-o-espelho/. Acesso em: 16 nov. 2019.

SANTA RITA, Bruno; FERRARI, Hamilton. **Renda de servidores eleva a média salarial no país, segundo o IBGE**. Correio Braziliense. 2019. Matéria jornalística. Disponível em: https://www.correiobraziliense.com.br/app/noticia/economia/2019/03/18/internas_economia,7 43567/renda-de-servidores-eleva-a-media-salarial-do-pais.shtml. Acesso em: 16 nov. 2019.

SANTOS, Fábio. **O que é uma Mesa Proprietária? Como funciona?** Meu day trade. 2019. Artigo em portal de conteúdo. Disponível em: https://meudaytrade.com/mesa-proprietaria/. Acesso em: 16 nov. 2019.

WOLWACZ, Alexandre. **Controle de Risco**. Porto Alegre: Leandro & Stormer. 2010. Publicação Eletrônica. Disponível em: https://docgo.net/view-doc.html?utm_source=controle-de-risco-alexandre-wolwacz-stormer-pdf&utm_campaign=download. Acesso em: 16 nov. 2019.